\documentclass[twocolumn,showpacs,preprintnumbers,amsmath,amssymb]{revtex4-1}
\usepackage{graphicx}
\usepackage{dcolumn}
\usepackage{bm}

\begin{document}

\title{Femtosecond pulses in a dense two-level medium: Spectral transformations, transient processes, and collisional dynamics}

\author{Denis V. Novitsky}
\email{dvnovitsky@tut.by} \affiliation{B.I. Stepanov Institute of
Physics, National Academy of Sciences of Belarus, \\
Nezavisimosti~Avenue~68, 220072 Minsk, Belarus}

\begin{abstract}
Propagation of ultrashort optical pulses in a dense resonant medium
is considered in the semiclassical limit. In our analysis, we place
emphasis on several main points. First, we study transformations of
spectra in the process of pulse propagation and interactions with
another pulse. The second point involves the transient processes
(including pulse compression) connected with self-induced
transparency soliton formation inside the medium. Finally, the third
aspect is the study of collisions of co- and counter-propagating
pulses in the medium. In the last case, the investigation of
symmetric and asymmetric collisions shows the possibility of
effectively controlling the parameters of transmitted radiation.
\end{abstract}


\maketitle

\section{Introduction}

A two-level medium, i.e. a medium composed of a collection of
two-level systems (atoms, molecules, etc.), is the basic
quantum-mechanical model in the theory of light-matter interaction.
It plays the same fundamental role as a harmonic oscillator in
classical physics. Therefore, the comprehensive study of this model
and its generalizations is one of the most important problems in
theoretical optics. The two-level model has allowed the description
of a number of well-known effects \cite{Allen, Kryukov} including
the effect of self-induced transparency \cite{McCall, Poluektov}.
Among the further generalizations of the model we single out the one
which takes into account the near dipole-dipole interactions between
the elements of the dense medium due to introduction into the
equations of the so-called local-field correction \cite{Bowd93,
Cren08}. In stationary regime, the presence of the local field
results in the remarkable effect of intrinsic optical bistability
\cite{Hopf, Ben-Aryeh, Friedberg, Hehlen}.

In the regime of optical pulses, the investigation of local-field
effects is linked with solitonic dynamics \cite{Bowd91, Afan02},
optical switching \cite{Cren92, Scalora}, etc. It was shown
\cite{Scalora} that, in the dense two-level medium, the scale of
population and field changes becomes so short that one cannot use
the common slowly varying envelope approximation in space (SVEAS) to
describe light pulse propagation. Therefore, one should use the full
(not truncated) Maxwell wave equation. This is the case in the
present paper where we consider one-dimensional problem of
ultrashort (femtosecond) light pulse propagation in a two-level
medium. Two remarks should be made here. First, as discussed in our
previous work \cite{Novit1}, the local-field effects depend strongly
on the duration of the pulse. For the femtosecond pulses considered
in this paper, these effects are negligible. However, the dynamics
of the pulse cannot be fully described by the usual self-induced
transparency (SIT) model. This fact leads us to the second remark.
In the dense two-level medium, calculations using the wave equation
without SVEAS show the slow attenuation of SIT solitons as was
predicted in Ref. \cite{Novit}. This and similar other effects
provide the rich dynamics of transient processes which are one of
the main objects of this investigation.

Another aim of our research is to study in detail the complex
dynamics of pulse collisions in the two-level medium. Interaction of
counter-propagating solitons was previously a research issue in both
the cases of two-level \cite{Afan, Shaw} and three-level media
\cite{Pusch, Park, Loiko}. In the appropriate section, we will
return to some of this works and discuss the further progress
achieved in this paper. Here, it is worth noting that we try to
trace step-by-step the changes in parameters of the pulses obtained
at relatively short distances due to the high density of the medium.
In addition, collisions of solitons and attendant effects were
theoretically and experimentally studied in different other systems
such as optical fibers \cite{Malomed}, photorefractive crystals
\cite{Krolik1, Krolik2}, and photonic crystals \cite{Petrovic,
Erkintalo}. In this paper we consider temporal solitons, while the
collisions between spatial ones were reported as well \cite{Cohen,
Petrovic}.

The structure of the paper corresponds to the gradual transition
from the relatively simple situations to the more complicated ones.
Section II is devoted to the case of single pulse propagation in a
dense two-level medium paying attention to the transformations of
light spectrum as it moves through the medium. In Section III we
consider interaction of co-propagating solitons, especially those
overtaking one another. Section IV is dedicated to the case of
counter-propagating pulses colliding in the medium. In particular,
we analyze the situation of the so-called symmetric collision there.
Finally, in Section V we study the asymmetric collisions between the
counter-propagating solitons.

\section{Single pulse propagation}

We start with the semiclassical Maxwell-Bloch system for population
difference $W$, microscopic polarization $R$, and electric field
amplitude $\Omega'=\Omega/\omega=(\mu/\hbar\omega)E$ (i.e.
normalized Rabi frequency) \cite{Bowd93, Cren96},
\begin{eqnarray}
\frac{dR}{d\tau}&=& i \Omega' W + i R (\delta+\epsilon W) - \gamma_2 R, \label{dPdtau} \\
\frac{dW}{d\tau}&=&2 i (\Omega'^* R - R^* \Omega') -
\gamma_1 (W-1), \label{dNdtau} \\
\frac{\partial^2 \Omega'}{\partial \xi^2}&-& \frac{\partial^2
\Omega'}{\partial \tau^2}+2 i \frac{\partial \Omega'}{\partial
\xi}+2 i \frac{\partial \Omega'}{\partial
\tau} \nonumber \\
&&=3 \epsilon \left(\frac{\partial^2 R}{\partial \tau^2}-2 i
\frac{\partial R}{\partial \tau}-R\right), \label{Maxdl}
\end{eqnarray}
where $\tau=\omega t$ and $\xi=kz$ are dimensionless arguments;
$\delta=\Delta\omega/\omega$ is the normalized detuning of the field
carrier (central) frequency $\omega$ from atomic resonance;
$\gamma_1=(\omega T_1)^{-1}$ and $\gamma_2=(\omega T_2)^{-1}$ are
the rates of longitudinal and transverse relaxation, respectively;
$k=\omega/c$ is the wavenumber, and $c$ is the light speed in
vacuum; $\epsilon=\omega_L/ \omega = 4 \pi \mu^2 C / 3 \hbar \omega$
is the normalized Lorentz frequency. Here we assume that the
background dielectric permittivity of the medium is unity (two-level
atoms in vacuum). Note that in Eq. (\ref{Maxdl}) we do not use
slowly-varying envelope approximation (SVEA) which cannot hold true
even for thin films of the medium \cite{Scalora}.

In this paper we consider propagation of ultrashort pulses with
Gaussian shape $\Omega=\Omega_p\exp(-t^2/2t_p^2)$ where $t_p$ is the
pulse duration. Amplitude of pulses is measured in the units of the
characteristic Rabi frequency $\Omega_0=\sqrt{2\pi}/2 t_p$, which
corresponds to the so-called $2 \pi$-pulse \cite{Novit}. For
calculations the values $T_1=1$ ns and $T_2=0.1$ ns are taken, so
that femtosecond pulses appear to be in the regime of coherent
interaction with the resonant medium. The spectra of pulses plotted
in this paper are obtained as the absolute values of the Fourier
transform of the corresponding field profiles. The spectra are
normalized on the peak value of the incident pulse spectrum which is
recognized as the unity. We use the following parameters of
calculations which hold true throughout the paper if the other is
not stated: $n_d=1$, $\omega_L=10^{11}$ s$^{-1}$, $\delta=0$,
$\lambda=0.5$ $\mu$m, $t_p=50$ fs.

\begin{figure}[t!]
\includegraphics[scale=0.9, clip=]{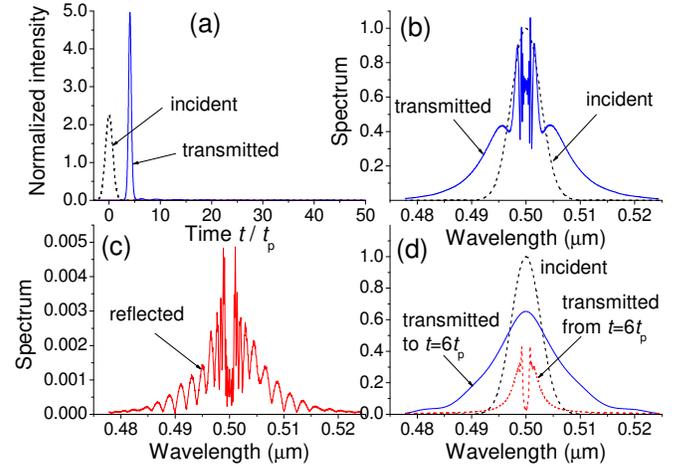}
\caption{\label{fig1} (Color online) (a) Intensity profile of the
pulse transmitted through the layer of thickness $L=100 \lambda$.
(b) and (c) Spectra of transmitted and reflected light respectively.
The spectra were calculated for the radiation appeared in the time
interval from $t=-3 t_p$ (start of the incident pulse) to $t=100
t_p$. (d) Spectra of transmitted light calculated for the radiation
appeared in the other time intervals (from $t=-3 t_p$ to $t=6 t_p$,
and from $t=6 t_p$ to $t=100 t_p$). The amplitude of the pulse
$\Omega_p=1.5 \Omega_0$. The local-field correction is not taken
into account.}
\end{figure}

In Fig. \ref{fig1} we demonstrate the results of propagation
simulation for the pulse with high peak amplitude ($\Omega_p=1.5
\Omega_0$). As it was shown in Ref. \cite{Novit}, an ultrashort
pulse in dense resonant medium experiences compression. The reason
for this is perhaps the effect of self-phase modulation reported in
Ref. \cite{Scalora}. The effect of compression is characterized by a
certain distance of optimal compression: After this distance the
pulse slowly attenuates due to dispersion. For the layer of
approximately optimal thickness, the main part of the pulse is
transmitted and compressed, while some part of its energy is
absorbed and then reemitted [Fig. \ref{fig1}(a)]. The spectrum of
the main part has the typical bell shape, while the reemitted light
has a dip at the resonant wavelength [Fig. \ref{fig1}(d)]. This can
be connected with effective reabsorption of low-intensity light at
the resonant frequency. At the same time the main, high-intensity
part of the pulse is almost not absorbed due to the self-induced
transparency (SIT) mechanism. The resulting spectrum of the
transmitted radiation is shown in Fig. \ref{fig1}(b): The two peaks
are situated at both sides of the central wavelength. The low-level
reflected light has a dip in its spectrum as well [Fig.
\ref{fig1}(c)].

\begin{figure}[t!]
\includegraphics[scale=0.9, clip=]{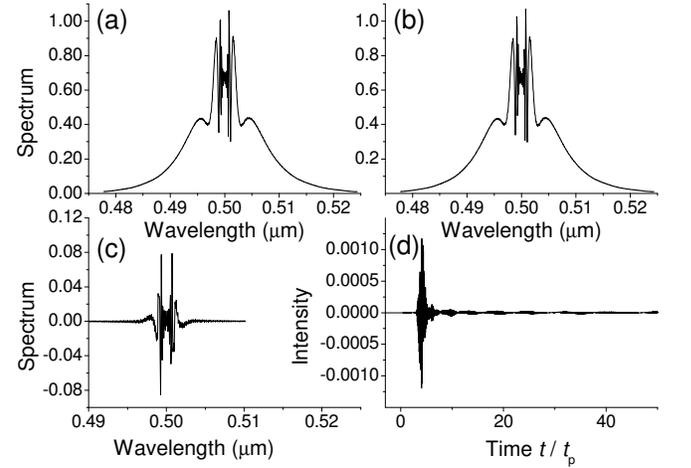}
\caption{\label{fig2} (a) and (b) Spectra of transmitted light in
the cases of absence and presence of local-field correction,
respectively. The spectra were calculated for the radiation appeared
in the time interval from $t=-3 t_p$ (start of the incident pulse)
to $t=100 t_p$. (c) Difference between the spectra shown in plots
(a) and (b). (d) Difference between intensity profiles of
transmitted light in the cases of absence and presence of
local-field correction, respectively. The amplitude of the pulse
$\Omega_p=1.5 \Omega_0$.}
\end{figure}

\begin{figure}[t!]
\includegraphics[scale=0.9, clip=]{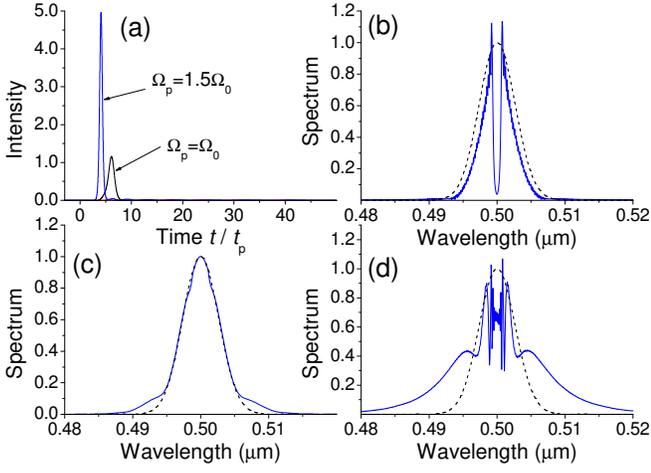}
\caption{\label{fig3} (Color online) (a) Intensity profiles of the
light transmitted through the layer of thickness $L=100 \lambda$ for
the pulses with different peak amplitudes. (b), (c), (d) Spectra of
transmitted light for the pulses with different peak amplitudes
($\Omega_p=0.5 \Omega_0$, $\Omega_0$ and $1.5 \Omega_0$,
respectively). The spectra were calculated for the radiation
appeared in the time interval from $t=-3 t_p$ (start of the incident
pulse) to $t=100 t_p$. The dashed line corresponds to the spectrum
of incident pulse. The local-field correction is taken into
account.}
\end{figure}

\begin{figure}[t!]
\includegraphics[scale=0.9, clip=]{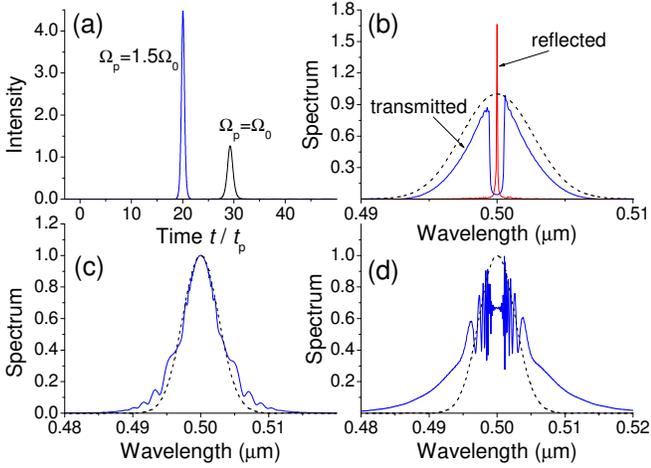}
\caption{\label{fig4} (Color online) (a) Intensity profiles of the
light transmitted through the layer of thickness $L=500 \lambda$ for
the pulses with different peak amplitudes. (b) Spectra of
transmitted and reflected light for the pulse with the peak
amplitude $\Omega_p=0.5 \Omega_0$. (c), (d) Spectra of transmitted
light for the pulses with the peak amplitudes $\Omega_p=\Omega_0$
and $1.5 \Omega_0$, respectively. The spectra were calculated for
the radiation appeared in the time interval from $t=-3 t_p$ (start
of the incident pulse) to (c), (d) $t=200 t_p$, (b) $t=600 t_p$. The
dashed line corresponds to the spectrum of incident pulse.}
\end{figure}

\begin{figure}[t!]
\includegraphics[scale=0.9, clip=]{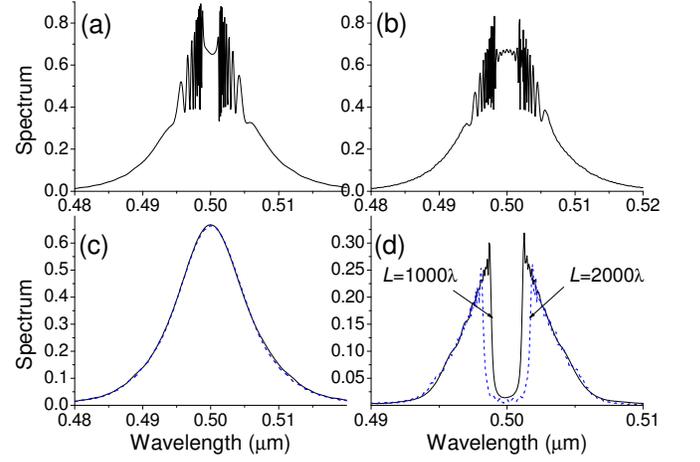}
\caption{\label{fig5} (Color online) Spectra of transmitted light
for the pulse with the initial peak amplitude $\Omega_p=1.5
\Omega_0$ after propagation of (a) $L=1000 \lambda$, (b) $L=2000
\lambda$. (c) Spectra of the main part of transmitted pulse and (d)
spectra of the "tail" for $L=1000 \lambda$ and $2000 \lambda$.}
\end{figure}

\begin{figure}[t!]
\includegraphics[scale=0.9, clip=]{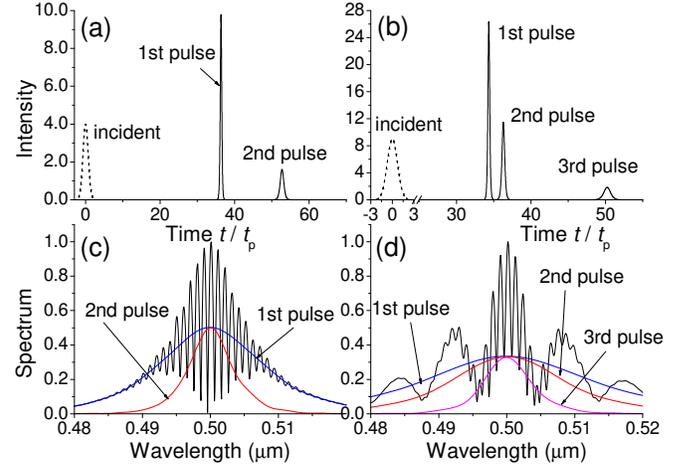}
\caption{\label{fig6} (Color online) (a), (b) Intensity profiles of
the light transmitted through the layer of thickness $L=1000
\lambda$ for the pulses with peak amplitudes $\Omega_p=2 \Omega_0$
and $3 \Omega_0$, respectively. (c), (d) Spectra of transmitted
light corresponding to the upper figures.}
\end{figure}

Figure \ref{fig1} was obtained in the absence of the local-field
correction (LFC), that is without the term $\epsilon W$ in Eq.
(\ref{dPdtau}). To evaluate the influence of local field, we
calculate the difference between the transmitted light spectra
obtained with and without LFC [Fig. \ref{fig2}(c)]. It is seen that
LFC leads to the significant change in spectrum (more than $8 \%$ of
the maximal value), while the intensity profile remains almost
unchanged [Fig. \ref{fig2}(d)] in accordance with the results
previously reported \cite{Novit1}. The difference between spectra is
mainly due to the late, reemitted radiation which interacts with the
medium for a time long enough to feel the local field. Nevertheless
the overall form of the spectrum is the same as one can ascertain
comparing Figs. \ref{fig2}(a) and (b). Therefore, hereinafter all
the results are obtained taking into account LFC.

Returning to Fig. \ref{fig1}, we should once more emphasize the role
of pulse "tail" in formation of the fine structure of the complex
spectrum of transmitted light. Note that the resulting spectrum is
not a simple superposition of the two others [see Fig.
\ref{fig1}(d)] because of interference term which should be taken
into account here. Figure \ref{fig3} shows the spectra for the
transmitted pulses with different initial peak amplitude $\Omega_p$.
The simplest case is observed for $\Omega_p=\Omega_0$, i.e. for the
$2 \pi$-pulse. Such a pulse is only slightly compressed and can
propagate for a long distance almost without change. Its spectrum is
generally coincident with that of the incident pulse [Fig.
\ref{fig3}(c)]. The pulses with amplitudes larger and smaller than
$\Omega_0$ demonstrate significant pulse transformations.
Low-intensity pulse (in our case $\Omega_p=0.5 \Omega_0$)
experiences strong absorption on the central frequency, so that the
output radiation has no any regular profile [it is not seen in Fig.
\ref{fig3}(a) due to its proximity to zero]. The overall energy
leaving the medium in this case is about half of the initial pulse
energy (for the calculational conditions denoted in the caption of
Fig. \ref{fig3}). This output energy is mainly concentrated in the
side bands of the spectrum as seen in Fig. \ref{fig3}(b). In the
case of high-intensity pulse ($\Omega_p=1.5 \Omega_0$), absorption
is not significant and almost all energy transmits through the
medium. However, its spectrum [Fig. \ref{fig3}(d)] implies that some
part of the radiation is transferred from the central wavelength to
the side bands.

The next step is to consider the transformations of spectra with
distance traveled by the pulses in the two-level medium (we increase
the distance by the factor of $5$, i.e. to $L=500 \lambda$). We find
that the spectrum of $2 \pi$-pulse gets more distorted as it
propagates inside the medium for a longer distance [Fig.
\ref{fig4}(c)]. In Fig. \ref{fig4}(b) one can see that the spectrum
of low-intensity transmitted pulse has the same form as previously.
In this figure the spectrum of reflected light is plotted as well.
Reflected radiation is concentrated mainly very close to the central
(resonant) frequency. In other words, the medium in this case can
serve as a very-narrow-bandwidth filter. However, the efficiency of
such a filter is not large because only a few percent of the
incident light is reflected.

The case $\Omega_p=1.5\Omega_0$ is of maximal interest to us due to
the strongly pronounced role of the pulse "tail". Comparison of
Figs. \ref{fig3}(d) and \ref{fig4}(d) shows that the side bands move
away from the resonant wavelength as the pulse propagates in the
medium. This is also proved in Fig. \ref{fig5}(a) and (b). This
transformation of spectrum is due to change in the spectrum of the
"tail": The central dip gets wider [Fig. \ref{fig5}(d)]. Note that
the main part of the pulse is characterized by the same spectrum
[Fig. \ref{fig5}(c)]. It remains almost unchanged because the main
part demonstrates invariant propagation in the regime of
self-induced transparency (SIT-soliton). However, the main pulse
very slowly attenuates though not so fast as was reported previously
\cite{Novit}. This is due to higher accuracy of our calculations in
the current investigation.

If the pulse amplitude is high enough ($\Omega_p \geq 2 \Omega_0$),
the splitting of the pulse into several solitons can be observed.
The examples of spectral transformations connected with this
splitting are shown in Fig. \ref{fig6}. Comparison between the
smooth spectra of separate solitons and the resulting jagged one
implies that the phase relations between frequency components plays
an important role. This interference is especially characteristic
for the pulse splitting into three components
($\Omega_p=3\Omega_0$). In this case [Fig. \ref{fig6}(d)], apart
from the central band, the spectrum has several pronounced and wide
side bands. It is also worth noting that the first soliton has
always the widest spectrum while every next one is more narrow. This
corresponds to the fact that the first pulse is more compressed than
any consecutive one.

\section{Co-propagating pulses}

\begin{figure*}[t!]
\includegraphics[scale=0.95, clip=]{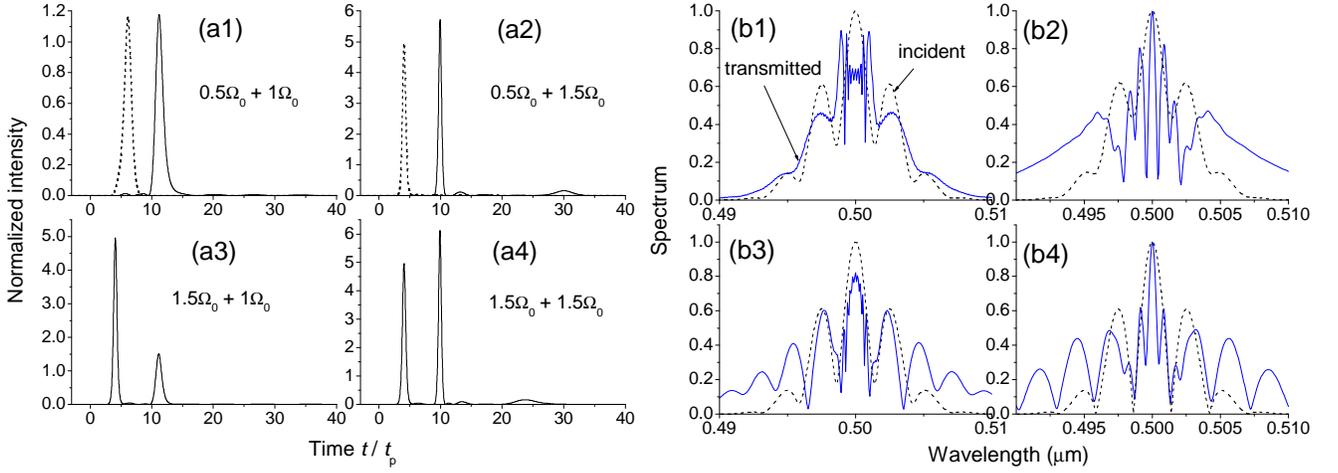}
\caption{\label{fig7} (Color online) (a) Intensity profiles and (b)
spectra of the light transmitted through the layer of thickness
$L=100 \lambda$ for the pairs of incident pulses. The numbers at (a)
and (b) correspond to different peak amplitudes of incident pairs:
(1) $0.5 \Omega_0 + \Omega_0$, (2) $0.5 \Omega_0 + 1.5\Omega_0$, (3)
$1.5 \Omega_0 + \Omega_0$, (4) $1.5 \Omega_0 + 1.5 \Omega_0$. The
time interval between the maxima of incident pulses is $6 t_p$. The
dotted line denotes (a) the transmitted single pulse with initial
amplitude $\Omega_0$ (1) and $1.5\Omega_0$ (2); (b) the spectra of
initial pair of pulses.}
\end{figure*}

\begin{figure*}[t!]
\includegraphics[scale=0.95, clip=]{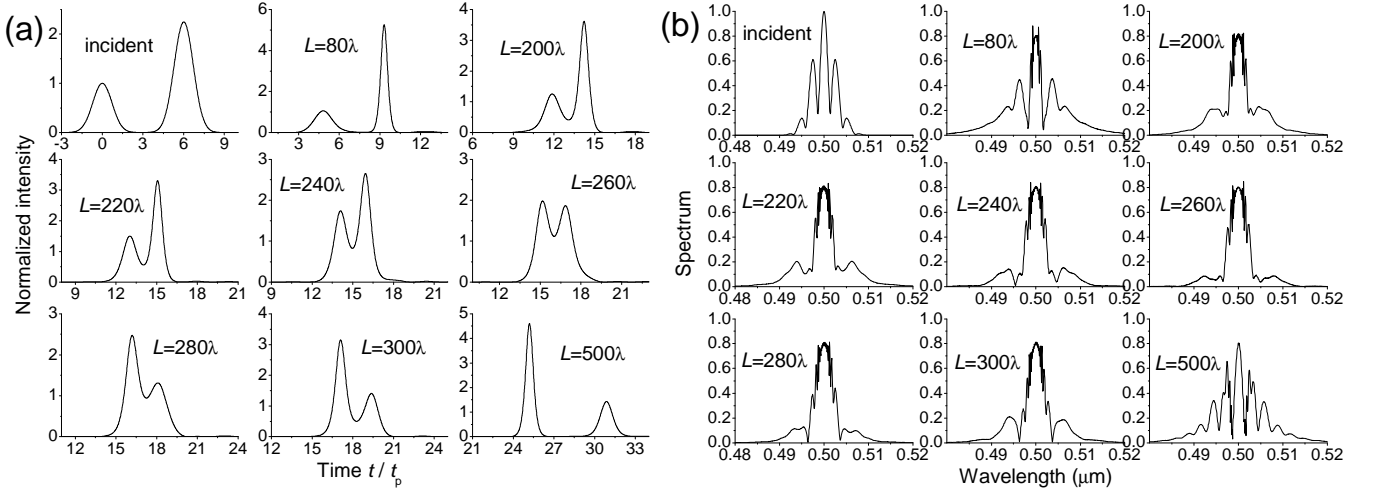}
\caption{\label{fig8} (a) Intensity profiles and (b) spectra of the
light transmitted through the layers of different thicknesses. The
calculations were performed for the pair of incident pulses
$\Omega_0 + 1.5 \Omega_0$. The time interval between the maxima of
incident pulses is $6 t_p$.}
\end{figure*}

\begin{figure}[t!]
\includegraphics[scale=0.95, clip=]{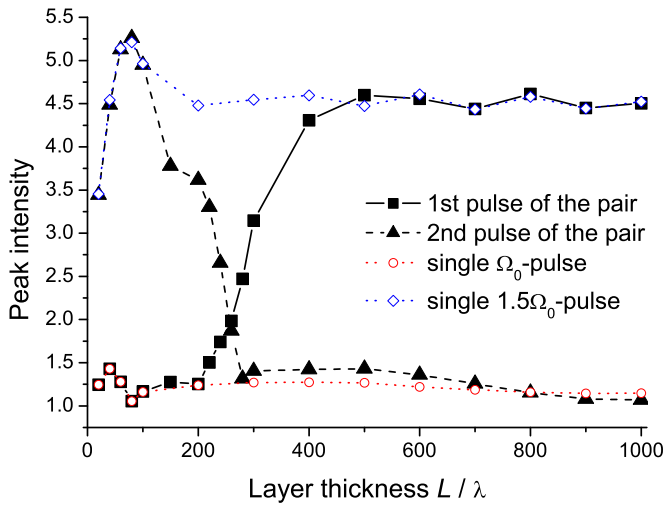}
\caption{\label{fig9} (Color online) Dependence of peak intensity of
the light transmitted through the layer on its thickness. The
calculations were performed for the pair of incident pulses
$\Omega_0 + 1.5 \Omega_0$ and for the single pulses $\Omega_0$ and
$1.5 \Omega_0$.}
\end{figure}

\begin{figure}[t!]
\includegraphics[scale=0.9, clip=]{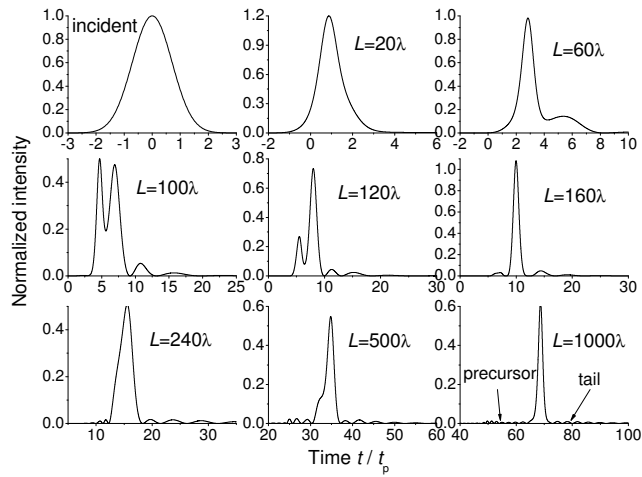}
\caption{\label{fig10} Intensity profiles of the light transmitted
through the layers of different thicknesses; the reflected light has
the identical envelope. The calculations were performed for the pair
of counter-propagating incident pulses with amplitude $\Omega_0$.}
\end{figure}

\begin{figure}[t!]
\includegraphics[scale=0.9, clip=]{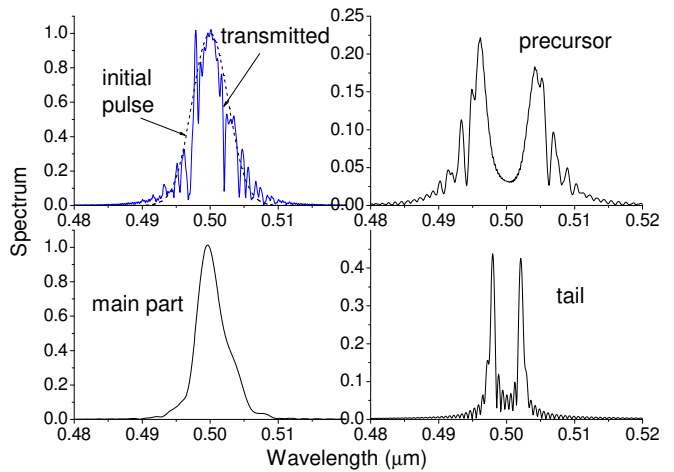}
\caption{\label{fig11} (Color online) Spectrum of light transmitted
through the layer of thickness $L=1000 \lambda$. The calculations
were performed for the pair of counter-propagating incident pulses
with amplitude $\Omega_0$.}
\end{figure}

\begin{figure}[t!]
\includegraphics[scale=0.9, clip=]{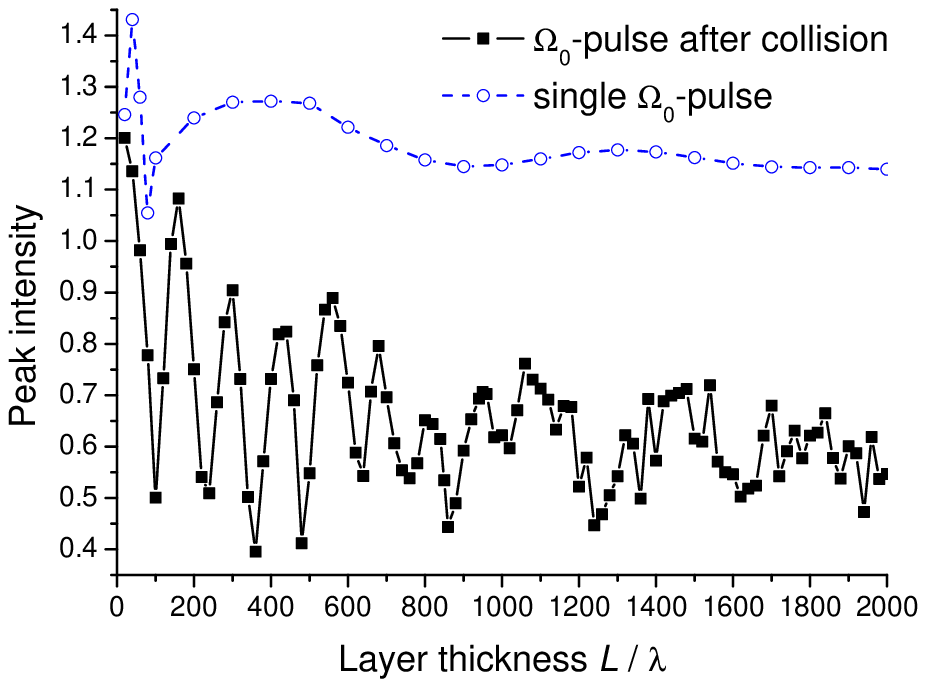}
\caption{\label{fig12} (Color online) Dependence of peak intensity
of the light transmitted through the layer on its thickness. The
calculations correspond to the case represented in Fig.
\ref{fig10}.}
\end{figure}

Now let us consider the process of interaction between two pulses
co-propagating in the dense two-level medium. Some preliminary
results were published previously \cite{Novit}. However, in this
paper we present extended and more accurate calculational results.
First, in Fig. \ref{fig7}(a) we show how the intensity of the second
pulse changes under the influence of the first one. The time
interval between the peaks of two incident co-propagating pulses is
taken to be $6 t_p$. It is seen that the first pulse loosing part of
its energy causes some increase in the intensity of the second one
in comparison with the case of the single pulse (this corresponds to
the dotted lines). This is especially evident in the case of Fig.
\ref{fig7}(a4). Figures \ref{fig7}(b) demonstrate the
transformations of spectra due to simultaneous compression and
interactions between the pulses. Note that the spectrum of the pair
$0.5 \Omega_0 + 1 \Omega_0$ [Fig. \ref{fig7}(b1)] is similar to the
spectrum of the single pulse with amplitude $1.5 \Omega_0$ [Fig.
\ref{fig4}(d)] and can be treated as composed from the spectra of
pulses with $0.5 \Omega_0$ and $\Omega_0$ [Fig. \ref{fig4}(b) and
(c)]. Other spectra are characterized by appearance of strong side
lobes and complex fine structure in the central part as seen in Fig.
\ref{fig7}(b2).

One particular case (the pair $1 \Omega_0 + 1.5 \Omega_0$) is worth
considering in detail here. This is because the second pulse of this
pair propagates in the medium faster than the first one as results
from Fig. \ref{fig4}(a). This means that, though the $2 \pi$-pulse
does not influence the second pulse at short distances, their
interaction should be getting stronger as they co-propagate in the
medium. The reason for this distinction in pulse velocity can be
easily understood on the basis of the well-known expression for the
stationary pulse \cite{Poluektov},
\begin{eqnarray}
\frac{1}{u}=\frac{1}{c}+\frac{\alpha t_p}{2},
\end{eqnarray}
where $u$ is the group velocity of the pulse, $c$ is the vacuum
light speed, $\alpha \sim C t_p$ is the extinction coefficient, $C$
is the concentration of two-level atoms, $t_p$ is the stationary
pulse duration. Here we are interested only in relative speed of
both pulses, so one can write
\begin{eqnarray}
\frac{c/u_1-1}{c/u_2-1}=\frac{C_1 t_{p1}^2}{C_2 t_{p2}^2},
\end{eqnarray}
or, taking into account that the peak intensity $I_p \sim 1/t_p^2$,
\begin{eqnarray}
\frac{c/u_1-1}{c/u_2-1}=\frac{C_1}{C_2}\frac{I_{p2}}{I_{p1}}.
\end{eqnarray}
This condition is perfectly satisfied as seen in Fig. \ref{fig4}(a)
(for the case $C_1=C_2$). The concentration dependence also holds
true as demonstrated by our careful examinations. This fact can be
treated as one more proof of validity of our calculational scheme.

Thus, we can study the process of collision of two pulses
co-propagating with different velocities. This case is demonstrated
in Fig. \ref{fig8}. It is seen that the second, more intensive pulse
overtakes the first one and, finally, at the length approximately
$L=250 \lambda$ they form a single pulse. This is indicated by the
dramatic simplification of the spectrum which retains only one,
central lobe. As the distance increases, this single pulse
disintegrates, and the intensive pulse leaves the other behind. This
fact corresponds to significant increase in complexity of the
spectrum. The results of Fig. \ref{fig8} also show that the
intensity of the single pulse after coalescence is intermediate
between the intensities of the pulses before the collision. The
change in the peak intensities of the pulses as they co-propagate in
the medium is depicted in Fig. \ref{fig9}. It is interesting to
compare them with the behavior of intensity of a single pulse. It is
seen that, after some transient process, the peak intensity of the
single pulse remains almost invariable. This corresponds to the
regime of solitonic propagation with the envelope perfectly
described by hyperbolic secant function (this means that $1.5
\Omega_0$-pulse with initial area of $3 \pi$ transforms into a
common $2 \pi$-soliton). However, the intensity of the pulse is not
strictly constant, it slightly fluctuates; so we cannot talk about
strict, \textit{mathematical} solitons but rather about realistic,
\textit{physical} ones. In addition, the amplitude of these
fluctuations tends to decrease with the propagation distance.
However, at larger distances the pulse slowly attenuates. For
example, the intensity of the pulse with initial amplitude $1.5
\Omega_0$ drops from approximately $4.5 \Omega_0^2$ (at $L=1000
\lambda$) to $3.75 \Omega_0^2$ (at $L=10000 \lambda$).

\begin{figure*}[t!]
\includegraphics[scale=0.95, clip=]{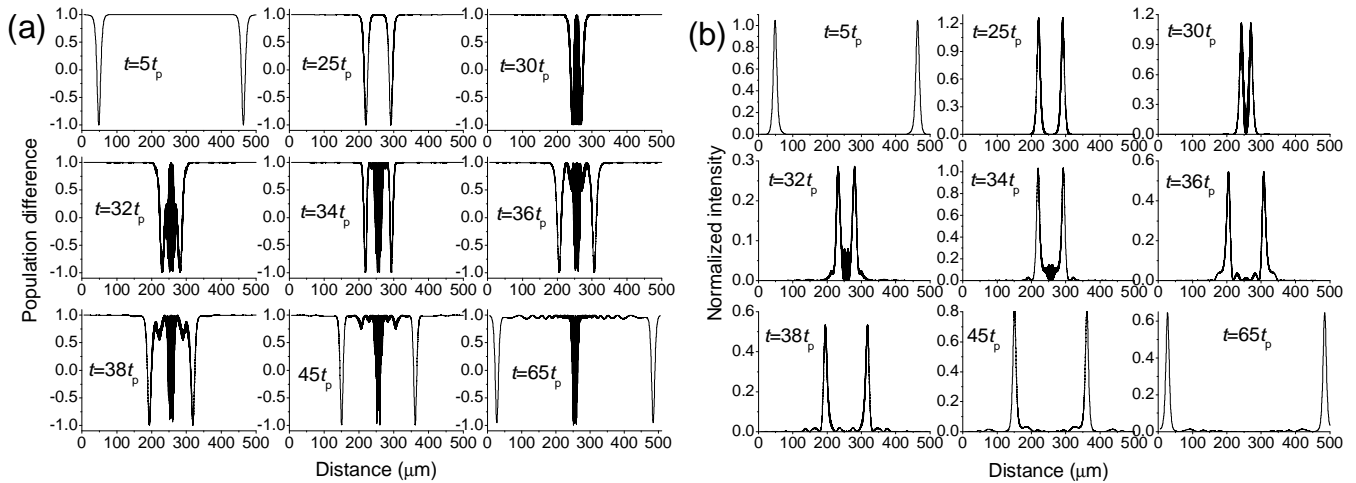}
\caption{\label{fig13} Distribution of (a) population difference and
(b) light intensity inside the layer $L=1000 \lambda=500$ $\mu$m at
different time points. The calculations were performed for the pair
of counter-propagating incident pulses with amplitude $\Omega_0$.}
\end{figure*}

The next question is connected with the nature of the transient
process: At first, the pulse is strongly compressed (on the distance
of about $100 \lambda$), and only then its envelope transforms to
the solitonic one (hyperbolic secant) which is connected with some
decrease in the pulse intensity. The initial compression of the
pulse is due to the self-phase modulation effect and is strongly
dependent on the form of the incident pulse. This compression can be
observed even for the pulses with initial envelope described by
hyperbolic secant. It is known that, at the same area (say, $2
\pi$), the $\textrm{sech}$-pulse is wider than the Gaussian one, so
that if the soliton is formed from the initial pulse, it would
obtain greater duration. But it is not the case: the soliton forms
from the compressed pulse. This compression can be characterized by
the length value, namely, the so-called distance of optimal
compression.

Returning to the collision of two pulses, we see (Fig. \ref{fig9})
that they strongly influence one another, but, after the change of
the propagation order, the distance between them increases and the
solitons are restored. One can say, that this figure demonstrates
the interaction between the already formed or just forming solitons
which are collision resistant.

\section{Counter-propagating pulses}

\begin{figure}[t!]
\includegraphics[scale=0.85, clip=]{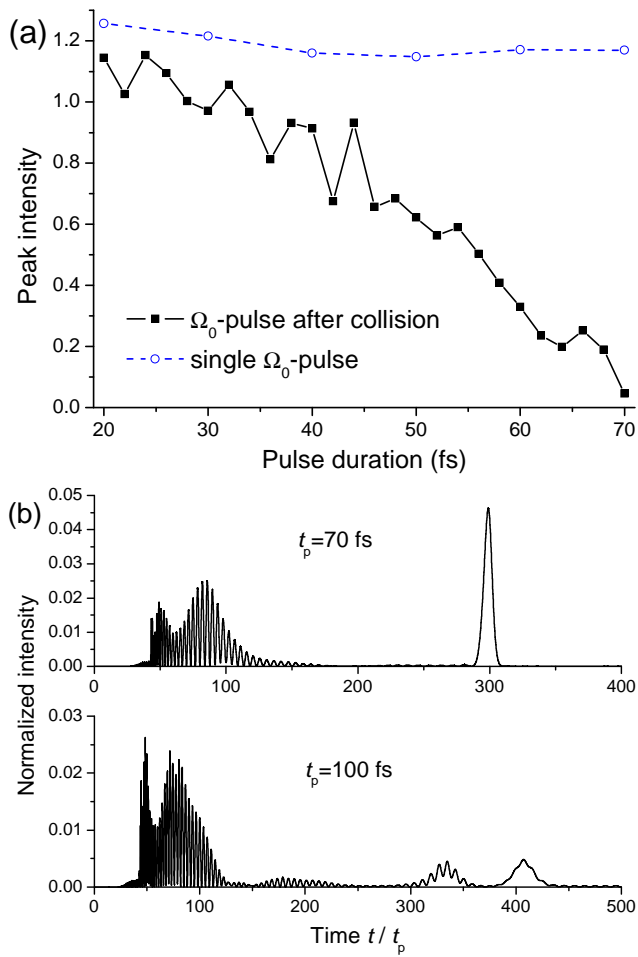}
\caption{\label{fig14} (Color online) (a) Dependence of peak
intensity of the light transmitted through the layer on the duration
of the initial $\Omega_0$-pulses. (b) Intensity of transmitted light
for initial pulse durations $t_p=70$ fs and $t_p=100$ fs. The
thickness of the layer is $L=1000 \lambda$.}
\end{figure}

From the previous section we know that the collision of
co-propagating pulses does not prevent formation of stationary
pulses with the same characteristics as in the single pulse case.
The collision of counter-propagating pulses has fundamentally
different results. The reasons were studied previously in Refs.
\cite{Afan, Shaw}. From mathematical point of view, two
counter-propagating pulses cannot provide a stationary (solitonic)
solution. Physically, this means that the absorption of some part of
energy of the pulses occurs in the region of collision. Further we
consider the dynamics of such collisions in detail.

We start from the collision of two Gaussian pulses with the
amplitude $\Omega_0$ entering the medium at the same instant of
time. In such symmetric geometry the light appearing from both ends
of the layer (we can conditionally call it transmitted and reflected
light) will be described by the same envelope. These envelopes are
shown in Fig. \ref{fig10} for different thicknesses of the layer. It
is seen that the result of the collision strongly depends on the
thickness $L$ and, as $L$ increases, the long precursor and tail
become apparent (in Ref. \cite{Shaw} only the tail was observed).
The spectra of the precursor and the tail are similar (Fig.
\ref{fig11}): they have a dip in the region of resonant wavelength,
while radiation at this central frequency is strongly absorbed.

To study this thickness dependence of the collision, we simulate the
process of interaction between the pulses in a wide range of
thicknesses $L$. The results of these calculations for the peak
intensity of the transmitted pulse are demonstrated in Fig.
\ref{fig12}. It is seen that at relatively small length there are
pronounced features which have a structure of something like
resonances. For larger thicknesses the amplitudes of these
resonances diminish, so that the peak intensity of the pulse tends
to the approximately constant value. This implies that the
resonances observed are connected with some transient process. This
process can be traced using distribution of population difference
and light intensity inside the medium at different instants of time
(Fig. \ref{fig13}). The thickness of the layer ($L=1000 \lambda$) is
large enough for $2 \pi$-solitons to be formed by the time of
collision. This stationary pulse is characterized by the typical
population difference profile (drop to full inversion and following
rise to the ground state). One can easily ascertain that, at the
location of collision, a strong absorption really occurs, so that
some part of light energy remains inside the medium even after the
pulses have gone away from each other. However, after collision,
pulse propagation is not simple: After reaching minimum at $t=32
t_p$, the pulse intensity grows ($t=34 t_p$), then goes down again
($t=36 t_p$), and so on. This process is obviously connected with
fluctuations depicted in Fig. \ref{fig9} and results in precursor
and tail formation. It is reasonable to suggest that the resonances
of Fig. \ref{fig12} are due to pulse exiting in different moments of
this transient process. But if the thickness of the layer is large,
these fluctuations become smaller and smaller, so that at the output
we have almost stationary pulse with the area $2 \pi$ again and
somewhat decreased intensity ($t=65 t_p$).

The next important feature of pulse collisions is their dependence
on the duration (or, equivalently, peak intensity) of the initial
pulses. This dependence is shown in Fig. \ref{fig14}(a) and is in
accordance with the results of Ref. \cite{Afan}: For short and
high-intensity $\Omega_0$-pulses, the influence of collision on
pulse propagation is not significant. On the contrary, long and
low-intensity pulses strongly interact with each other and, if the
duration is large enough, the main pulse (soliton) disappears
leaving only a precursor and a tail [see Fig. \ref{fig14}(b) for the
pulses with $t_p=100$ fs]. Note that the area of the main pulse ($2
\pi$) is conserved even in the case of dramatic decrease of its peak
intensity. For example, the pulse corresponding to $t_p=70$ fs in
Fig. \ref{fig14}(b) can be easily fitted by a hyperbolic secant
function. This means that such low-intensity solitons undergo
substantial broadening.

\section{Asymmetric collisions}

\begin{figure}[t!]
\includegraphics[scale=0.85, clip=]{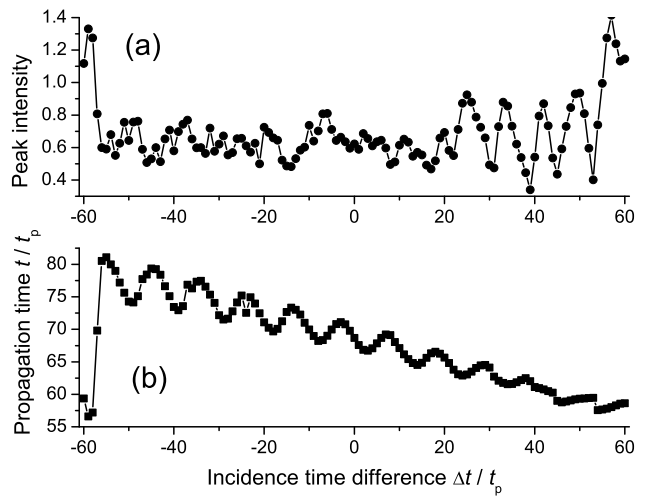}
\caption{\label{fig15} (a) Dependence of peak intensity of the light
transmitted through the layer on the incidence time difference
between the two initial counter-propagating $\Omega_0$-pulses. (b)
Corresponding behavior of propagation time calculated for the main
pulse (soliton). The thickness of the layer is $L=1000 \lambda$.}
\end{figure}

\begin{figure}[t!]
\includegraphics[scale=0.85, clip=]{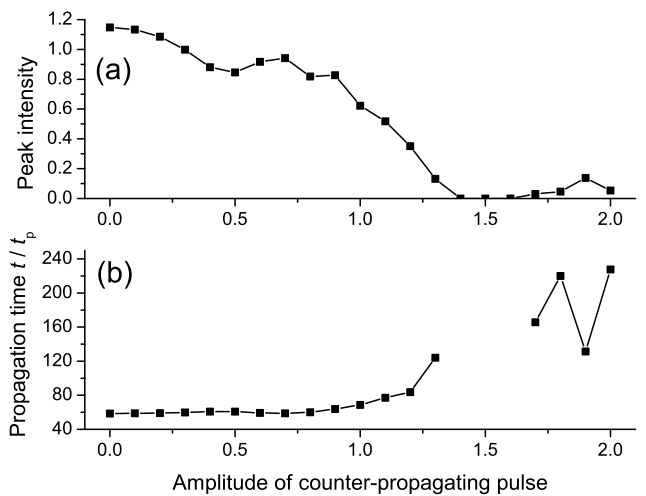}
\caption{\label{fig16} (a) Dependence of peak intensity of the light
transmitted through the layer (for initial $\Omega_0$-pulse) on the
amplitude of the counter-propagating pulse. (b) Corresponding
behavior of propagation time calculated for the main pulse
(soliton). The thickness of the layer is $L=1000 \lambda$.}
\end{figure}

\begin{figure*}[t!]
\includegraphics[scale=0.95, clip=]{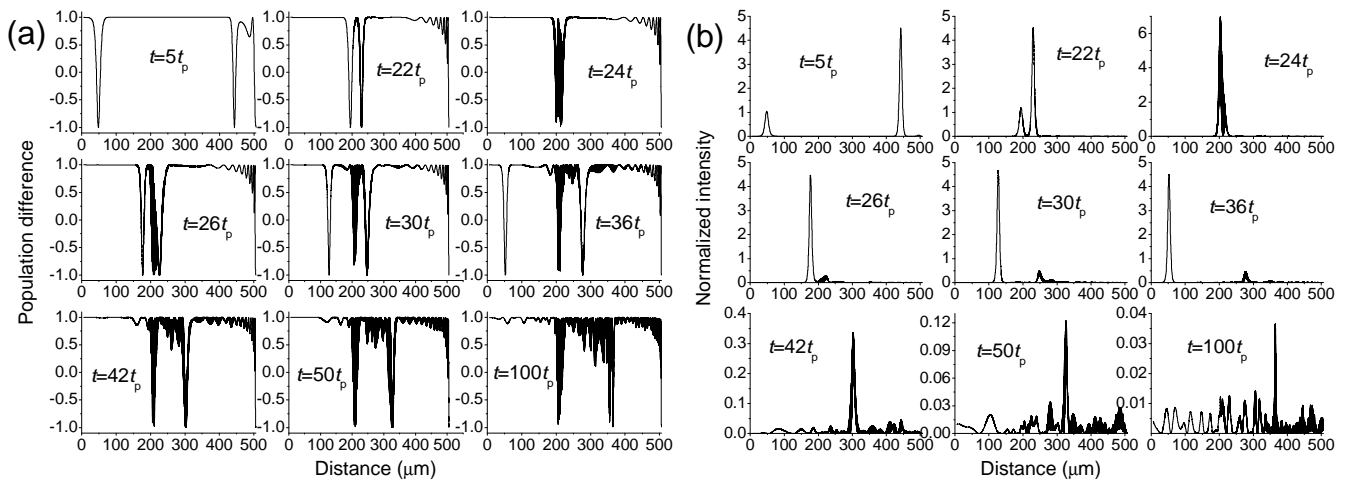}
\caption{\label{fig17} Distribution of (a) population difference and
(b) light intensity inside the layer $L=1000 \lambda$ at different
time points. The calculations were performed for the pair of
counter-propagating incident pulses: one with amplitude $\Omega_0$
and the other with $1.5 \Omega_0$.}
\end{figure*}

In previous section we considered the process of collision of the
two identical counter-propagating pulses entering the medium at the
same instant of time. Such geometry of the problem can be referred
to as the symmetric one. Further we discuss the case of asymmetric
collisions when the incidence time or intensity is different for
both initial pulses.

Figure \ref{fig15} shows the results of collision of the pulse with
the amplitude $\Omega_0$ with the identical pulse which enters the
medium at different time. Negative values of the incidence time
difference $\Delta t$ means that the counter-propagating pulse is
launched earlier than those propagating in the forward direction
($\Delta t=0$ corresponds to the symmetric situation considered in
the previous section). In the case of $\Delta t < 0$, the collision
occurs closer to the input of the forward-propagating pulse, so that
it has much time to form a stationary pulse after the collision.
Therefore, the fluctuations of the peak intensity [Fig.
\ref{fig15}(a)] are less pronounced in this case than for $\Delta t
> 0$. On both edges of this dependence (that is, for large absolute
values of $\Delta t$) one obtains the trivial case of the single
pulse propagation. The transition to this extreme regime has
jump-like character on both sides of the peak intensity dependence.
This symmetry does not take place for the time of pulse transmission
through the layer [Fig. \ref{fig15}(b)]. For large negative $\Delta
t$ the abrupt increase of time propagation occurs due to formation
of the low-intensity (and, hence, slow) soliton near the very input.
As the counter-propagating pulse enters the medium later and later,
the fast initial pulse has enough time to pass a large fraction of
distance before the collision happens. As a result, the propagation
time decrease gradually as seen in Fig. \ref{fig15}(b). Thus, the
collisional dynamics discussed imply the possibility of controlling
such pulse parameters as its peak intensity and transmission time by
proper choice of incidence time of the counter-propagating pulse.

Another approach to controlling the parameters of the transmitted
pulse is to use the counter-propagating pulses of different
intensities. In Fig. \ref{fig16} one can see the change in peak
intensity of $\Omega_0$-pulse as a function of the amplitude of the
counter-propagating (controlling) one. As this amplitude grows, the
peak intensity of the transmitted forward-propagating pulse
decreases as well as its propagation velocity. Finally, in the
region of amplitudes near $1.5 \Omega_0$, the transmitted solitonic
pulse is absent \textit{per se}. We have only the precursor and the
tail. For larger amplitudes of controlling pulse, the transmitted
one appears again, of course, with larger retardation. However,
collisional dynamics in this case are even more complex due to the
effects of pulse splitting. Therefore, we will not consider the
details of these multi-pulse interactions in the present
investigation.

Instead, we turn to the case of complete soliton disappearance at
the output of the medium when the amplitude of the controlling pulse
equals $1.5 \Omega_0$. The dynamics of level population difference
and light intensity inside the medium depicted in Fig. \ref{fig17}
imply that the primary importance in this effect is likely to be
connected with the residual excitation of the medium after passing
the controlling pulse. Since its initial area is $3 \pi$ (amplitude
$1.5 \Omega_0$), before the $2 \pi$-soliton formes, some part of the
energy is absorbed giving rise to the pronounced tail. It is seen
that the collision leads to the significant loss of energy of the
forward-propagating pulse while the controlling (high-intensity)
pulse remains almost unchanged. Moving farther, the low-intensive
pulse meets the tail of the counter-propagating one and the residual
excitation and, finally, undergoes ultimate absorption in the
medium. This absorbed energy gives fluorescent radiation in the long
run as was stated in Ref. \cite{Shaw}.

\section{Conclusion}

In this paper, we have considered the ultrashort (femtosecond) pulse
interaction with a dense collection of two-level atoms. In other
words, the light-medium interaction was investigated in the coherent
regime when incoherent phenomenological relaxation can be neglected.
Our study was conducted in semiclassical approximation, which is
valid as we are not interested in the detailed description of the
processes of spontaneous radiation \cite{Allen}. In addition,
semiclassical approach allows us to correctly take into account the
local field correction \cite{Cren08}. Our calculations were
performed for two-level atoms located in vacuum which corresponds to
dense gaseous media. In order to consider solid-state systems
(two-level atoms embedded in a dielectric, for example, quantum dots
in semiconductor or glassy environment), one has to take into
account the so-called local-field enhancement factor, as it was done
in stationary regime \cite{Cren96a, Novit2}. However, if the
background dielectric is not absorptive, behavior of optical pulses
reported in this paper is expected to be still valid for this more
complex case (with the corresponding renormalization of the electric
field).

In conclusion, we should discuss the prospects of pulse collision
dynamics studied in the paper. Our accurate calculations show the
self-healing effect for the co-propagating solitons but this is not
the case for the counter-propagating pulses. Together with the
transient processes of soliton formation, the strong interaction of
the counter-propagating pulses provides for the rich dynamics of
population difference and light intensity. The proper choice of the
parameters of the colliding pulses allows one to effectively control
the characteristics of light at the output of the medium. For
example, one can obtain almost entire absorption of the soliton
whose energy can be stored inside the medium, at least, for the
relaxation time. The conditions of controlling pulse intensity by
choosing the intensity or incidence time of the counter-propagating
one can be considered as a basis of the peculiar logic gates or
other elements for which the possibility of realization is still to
be studied.

\acknowledgements{The author is grateful to Alexander Kalinovsky for
help in performing numerical simulations.}

\end{document}